\documentstyle[12pt]{article}
\def\beq{\begin{equation}} 
\def\eeq{\end{equation}} 
\def\bea{\begin{eqnarray}} 
\def\eea{\end{eqnarray}} 
\def\bq{\begin{quote}} 
\def\eq{\end{quote}}

\def\bq{\begin{quote}} 
\def\eq{\end{quote}}

\def\un{\underline}
\def\ov{\overline}
\def\154{\frac{15}{4}}
\def\153{\frac{15}{3}}
\def\32{\frac{3}{2}}
\def\254{\frac{25}{4}}
\begin{document}
\newcommand{\gl}{\lambda}
\thispagestyle{empty} 
\begin{flushright} 
IOA.01/96\\
NTUA 54-96
 
\end{flushright} 
\begin{center} 
{\bf Supercompositeness From  Superstrings}\\

\vspace*{1cm} 
{\bf P. Dimopoulos$^{(a)(1)}$, 
     G.K. Leontaris$^{(b)(2)}$ and
     N.D. Tracas$^{(a)(3)}$} \\
\vspace*{0.3cm} 
{\it $^{(a)}$Physics Department},
{\it National Technical University} \\
{\it GR-157 80 Zografou, Athens, Greece} \\ 

\vspace*{0.6cm} 
{\it $^{(b)}$ Theoretical Physics Division},
{\it Ioannina University} \\
{\it GR-451 10 Ioannina, Greece} \\ 

\vspace*{0.5cm} 
{\bf ABSTRACT} \\ 
\end{center} 
\noindent 
String Unified Models based on the $k=1$ level of the Kac-Moody
Algebra, predict the existence of ``exotic'' new states which carry
fractional electric charges. We analyse the possibility of
considering these ``exotics'' as preonic matter which can be used
to form the families and the gauge group breaking higgs scalars.
It is proposed that such a formation may occur provided that these
states transform non-trivially under a non-Abelian gauge group
with a relatively large rank in order to confine them at a 
sufficiently large scale. Such a situation is natural in string
derived unified models, since the role of the confining group can 
be played by  (part of) the Hidden  symmetry. 
 As an example,  we present a string derived toy model based on 
the $SU(4)\times SU(2)_L \times SU(2)_R$ Pati-Salam gauge group.

\noindent
\begin{flushleft} 
IOA-01/96 \\
NTUA 54-96

\end{flushleft} 
\vfill
\hrule
$^{(1)}$pdimop@zeus.central.ntua.gr\\
$^{(2)}$leonta@cc.uoi.gr\\
$^{(3)}$ntrac@zeus.central.ntua.gr
\eject
 
\setcounter{page}{1}

 One of the most attractive and interesting issues in strings, 
is the construction\cite{mb} of realistic models which are
consistent with  the low energy phenomenology. Most of the attempts 
in string model  building
\cite{aehn,..,alr,af,cchl,l},
have been concentrated in constructions of string models based
on level-one (k=1) Kac-Moody algebras. 
In the search for the realistic string derived
models, two main obstacles have appeared:  \\
(i) Unified models based on these constructions do not contain fields
in the adjoint or higher representations. Therefore, traditional 
Grand Unified Theories (GUTs), like $SU(5)$ and $SO(10)$  could not 
break down to the Standard Model. Attempts to overcome this difficulty 
led to  the construction of models where the gauge group needs only small
Higgs representations to break\cite{aehn,alr}.\\
(ii) The appearance of fractionally charged states, other than the 
ordinary Quarks, is unavoidable\cite{fc} in the k=1 Kac-Moody
constructions. Such states,  unless they become massive at the String 
scale, they usually create problems in the low energy effective theory.
Indeed, the lightest fractionally charged particle is expected to be
stable. In particular, if its mass lies in the TeV region, then
the estimation of its relic abundances
\cite{ra} 
contradicts the upper
experimental bounds by several orders of magnitude. It has been
proposed that this problem  can in principle be solved by constructing 
models containing a hidden gauge group which  becomes strong at an
intermediate scale to confine the fractional charges into bound
states\cite{aehn,eln}.
 
In this work, we would like to explore an alternative scenario:
Since the fractionally charged states are generic\cite{fc} in k=1
level, it might be possible that in particular string  models they
could play the role of some {\em preonic} matter superfields which
transform non-trivially under some {\em Hidden} gauge group. This
{\em Hidden} group could very well play the role of the confining
gauge group of the preonic fields into composite states which could
be the representations containing the ordinary Quarks and Leptons. 

Models with composite Quarks and Leptons have already been introduced 
by many people\cite{th,revs,ven,pa,oth} the last two decades,
either in the context of the Standard gauge group of Electroweak 
interactions{\footnote{Compositeness may also be combined with
Technicolor and Extended Technicolor Theories\cite{dimsus}
to produce interactions which may create dynamically 
the fermion masses.}}
or within Grand  Unified schemes. Both scenarios are
well motivated in the context of Superstring k=1 constructions.
Indeed, if we insist on the economy of the models derived from
the String, we would feel unhappy with a large variety of  
representations left in the light spectrum of the effective
field theory, even if we manage with a judicious choice of the
parameters of the theory (moduli, flatness conditions etc) to 
make them massive at some intermediate scale $M_C$. Instead,
it  would appear more natural to derive an effective field 
theory with a relatively small number of representations.

In what follows, motivated by the appearance of such exotic 
states in string constructions, we will concentrate on a 
particular gauge group which leads to a viable low energy
scenario. In particular we will explore the $SU(4)\times
SU(2)_L\times SU(2)_R$ Pati-Salam symmetry assuming the
existence of representations carrying fractions of the known
electric charges possessed by the ordinary Quarks and Leptons. 
In this work, we will not discuss long standing problems 
which arise when trying to implement the idea of compositeness.
Relevant discussions about the problems appearing in the
various approaches to the compositeness may be found in
the literature\cite{th,revs,pec}. 

 In the
free fermionic four dimensional constructions
\cite{mb}, 
in principle it is possible to choose boundary conditions on 
the world sheet fermions of the basis vectors of a particular
model and project out all the integrally charged representations
{\footnote{In the $SU(4)\times SU(2)_L\times SU(2)_R$ model
this is rather obvious. Indeed the charge operator is a combination
of diagonal generators of all the group factors. Therefore, in
the fermionic language for example, we may extend the basis
by adding ``phase-vectors'' until all representations transform
non-trivially under only one of the non-Abelian groups. 
Many of these representations do not belong
to any known ``GUT'' multiplets and possess charges which are fractions
of those of the ordinary fermions.}}.
The fermion families, will appear then at an intermediate scale
as composite fields of the preonic representations. Therefore, 
it would be natural to ask, if any phenomenologically viable
preonic model could arise from the Plank scale physics.

 The necessary fields of the minimal supersymmetric version 
together with their transformation properties under the PS-gauge 
symmetry are shown in the table
\begin{eqnarray}
\begin{array}{llll}
F_L=(4,2,1) ;&{\bar F_R}=({\bar 4},1,2) ;& 
H=(4,1,2) ;&{\bar H}=({\bar 4},1,2) ;             \nonumber\\
h=(1,2,2) ;&D=(6,1,1);&\phi_{m,0}=(1,1,1), &m=1,2,3 \nonumber
\label{spec}
\end{array}
\end{eqnarray}
 The superpotential of the model can be written as follows
\bea
\cal W &=&\lambda_1F_L\bar F_Rh+\lambda_2\bar F_RH\phi_i
+\lambda_3\phi_0hh +\lambda_4\phi^3\nonumber\\
&{}&+\lambda_5HDD+\lambda_6\bar H\bar HD+\lambda_7\bar F_R\bar F_RD+
\lambda_8F_LF_LD\label{supe}
\eea

The  superpotential (\ref{supe}) includes  trilinear terms with states
arising only from the decomposition of the ordinary irreducible
representations (irreps) of the $SO(10)$ theory. At k=1  Kac-Moody level
 in particular, all irreps appearing in the theory are smaller than 
the adjoint. For the model under consideration for example, the 
possible representations under $SU(4)\times SU(2)_L\times SU(2)_R$
arise from the  decompositions of $\un{16}$, $\un{\ov{16}}$ and $\un{10}$ 
of SO(10).
In string constructions, however, the case is more complicated.
In fact, in this particular model the effective theory gauge symmetry
is based on  a product of non-Abelian groups rather than on a single
unified one. In the fermionic constructions for example, the model 
is constructed from a set of vectors whose components are phases picked  
 up by the world-sheet fermions when parallel transported around
non-contractable loops. The massless states of the theory are those
surviving the projections of the various vectors onto the others.  As a
result, in addition to the above states, new representations may arise
which are singlets under all but one of the non-Abelian factors of the
symmetry of the model.   Thus in addition to $(4,2,1),(\bar{4},1,2)$ one
may get the ``exotics'' $(4,1,1),(\bar{4},1,1)$, while together with
$(1,2,2)$ one also obtains $(1,2,1)$ and $(1,1,2)$. Of course such
representations are not present in the ordinary $SO(10)$ irrep
decompositions.

There are two ways of handling these states:\\
{\it i)} 
One can redefine the charge operator\cite{fc,alr,dfm}.
Indeed, in the usual string
constructions the resulting ``observable'' gauge symmetry is 
accompanied by  ``hidden'' gauge groups and a rather large
number of U(1) factors. Most of the fields discussed above
carry non-zero charges under the surplus U(1)'s. One then could
extend the charge operator by including one or more of these
U(1)'s. Such cases have been discussed in the literature but
they usually lead to the wrong predictions for the weak mixing
angle. \\
{\it ii)} As a second possibility we consider the case discussed
here where the string model predicts only the ``exotic'' states
discussed above, with non-trivial  transformation properties under      
part of the hidden gauge group. String toy models with such properties
can be easily constructed\cite{alp}.

With the form of the minimal theory in mind, let us now attempt
to derive it considering only {\em preonic} fields, assuming that the
ordinary superfields are not present in the original theory.
As has already been discussed, we assume that the symmetry
of the observable sector is based on the  gauge group 
$SU(4)\times SU(2)_L\times SU(2)_R$, while the fields
belong also to some $ N ({\bar N}) $-dimensional
representation of a {\em Hidden} gauge group. 
The fractionally charged states which appear in the string spectrum
of these models, in the most general case, are of the following types
\begin{eqnarray} 
\begin{array}{llllllll}  
K_{j} & = & (4,1,1)_{jN} & {}
& K^c_{n} & =&(\bar 4,1,1)_{n{N}}  \\
\bar K_{n} & = & (4,1,1)_{n{\bar N}}& {}
& \bar K^c_{j} & = & (\bar 4,1,1)_{j\bar N} \\
\alpha_{Li} & = & (1,2,1)_{iN}& {} 
& \alpha_{Rm} & = &(1,1,2)_{mN} \\ 
\bar\alpha_{Lm} & = & (1,2,1)_{m \bar {N}}
& {} & \bar \alpha_{Ri} & = & (1,1,2)_{i\bar {N}} \\
\end{array}
\end{eqnarray} 
Let us explain our notations in the above fields. The numbers 
in the parentheses, as usually, refer to the transformation properties
of  the various {\em preonic} fields under the observable gauge 
symmetry of the model. The indices $i,j,m,n$ refer to the number
of the corresponding representations and run from $1$ to $\cal I,\cal
J, \cal M,\cal N$ respectively. Care has been taken, so as ($4/\bar 4$)
as well as ($N/\bar N$) representations appear in pairs to ensure
that the theory is anomaly free. The index $N(\bar N)$ in each of
the above representations refers to its transformation property
under the {\em Hidden} gauge group 
\footnote {Sextets under $SU(4)$ can also arise at this level.}.

We should note here, that in realistic string constructions, 
the fields might also carry extra $U(1)$-charges while the 
{\em Hidden} gauge group is not always simple. However, in
 order to make the subsequent  analysis simple and
model independent, we consider only a simple $SU(N)$-{\em Hidden}
gauge group. The existence of the $U(1)$ factors would have the
obvious implication of reducing the possible gauge invariant trilinear
and higher order Yukawa terms of the superpotential.

Now, if we define the charge operator in the usual sense
\beq
{\cal Q}= \frac{1}{6}T_{15}+\frac{1}{2}T_L+\frac{1}{2}T_R
\eeq
where $T_{15}=diagonal (1,1,1-3)$, and $T_{L,R}=diagonal (1,-1)$,
it is clear that all the above fields carry charges which are
fractions of those of ordinary Quarks and Leptons.
Under the {\em Hidden } gauge group, they form composite states
at some intermediate scale $M_U< M_C < M_{Pl}$, which may be
identified with the ordinary superfields of Quarks and Leptons.
The possible composite states created from the above  {\em preonic}
fields are listed bellow

\begin{center}{\bf  Table I}\end{center}

\bea
\begin{array}{ccccccccccc}
(F_L)_{jm}&=&K_j\bar \alpha_{Lm}&=&(4,2,1)& {;} 
&(F_R)_{ji}&=&K_j\bar\alpha_{Ri}&=&(4,1,2)\\
(F_L)_{ni}&=&\bar K_n\alpha_{Li}&=&(4,2,1)& {;}
&(F_R)_{nm}&=&\bar K_n\alpha_{Rm}&=&(4,1,2)\\
(\bar F_L)_{nm}&=&K_n^c\bar \alpha_{Lm}&=&(\bar 4,2,1)& {;}
&(\bar F_R)_{ni}&=&K_n^c\bar \alpha_{Ri}&=&(\bar 4,1,2)\\
(\bar F_L)_{ij}&=&\bar K_j^c\alpha_{Li}&=&(\bar 4,2,1)& {;}
&(\bar F_R)_{jm}&=&\bar K_j^c\alpha_{Rm}&=&(\bar 4,1,2)\\
D_{jn}&=& K_j\bar K_n&=&(6,1,1)& {;}
&D_{nj}&=& K_n^c\bar K_j^c&=&(6,1,1)\\
T_{jn}&=& K_j\bar K_n&=&(10,1,1)& {;}
&T_{nj}&=& K_n^c\bar K_j^c&=&(10,1,1)\\
\Sigma_{jj}&=& K_j\bar K_j^c&=&(15,1,1)& {;}
&\bar {\Sigma}_{nn}&=& K_n^c\bar K_n&=&(15,1,1)\\
\Phi_{jj}&=& K_j\bar K_j^c&=&(1,1,1)& {;}
&\bar {\Phi}_{nn}&=& K_n^c\bar K_n&=&(1,1,1)\\
h_{ii}&=& \alpha_{Li}\bar \alpha_{Ri}&=&(1,2,2)& {;}
&h_{mm}&=&\alpha_{Rm}\bar \alpha_{Lm}&=&(1,2,2)\\
\Delta_{Lim}&=& \alpha_{Li}\bar \alpha_{Lm}&=&(1,3,1)& {;}
&\Delta_{Rmi}&=&\alpha_{Rm}\bar \alpha_{Ri}&=&(1,1,3)\\
\Phi'_{im}&=& \alpha_{Li}\bar \alpha_{Lm}&=&(1,1,1)& {;}
&\bar {\Phi'}_{mi}&=&\alpha_{Rm}\bar \alpha_{Ri}&=&(1,1,1)
\end{array}\nonumber
\eea

We observe that all the fields of the superpotential in Eq.(\ref {supe})
are present in the above table. The indices $\{i,j,m,n\}$ indicate
the multiplicity of each representation.  Thus, in the general case
considered above, one~obtains
$\cal J \cal M +\cal I \cal N$ left
handed fields $F_L$ and an equal number of right handed
representations $\bar F_R$. These representations are going to
accommodate the known fermion families of quarks and leptons and 
their superpartners. Note however, that the above are accompanied by 
$\cal I \cal J +\cal M \cal N$ ``anti-left''  $\bar F_L$,
and  ``anti-right'' $ F_R$  fields. The rest of the composite spectrum
includes  ${\cal I}^2 +{\cal M}^2$ higgses in (1,2,2), $2\cal J \cal N$
(6,1,1)-sextets and an equal number of (10,1,1) irreps.
The new feature is the appearance of ${\cal J}^2+  {\cal N}^2$ adjoint
representations (15,1,1) of SU(4) and ${\cal I}{\cal M}$ (1,3,1) 
and (1,1,3) as well as 
$2{\cal I}{\cal M} + {\cal J}^2+  {\cal N}^2$  neutral singlets.

In order to avoid the appearance of ``antifamilies'' in the light
spectrum, they should combine with equal number of families and
receive mass at a high scale. If we demand $r$ generations to
remain in the light spectrum, then we should have $\#F_L-\#\bar F_L=r$,
and an equal number of right partners. This requirement leads to
the equation $({\cal M}-{\cal I})\times ({\cal J}-{\cal N})=r$, 
which is satisfied for various choices of
${\cal I} ,{\cal J},{\cal M} $, and ${\cal N} $.

Thus, let us distinguish some simplified cases:

$\bullet \;{\cal M}=0.$ In this case, the above requirements for $r=3$
lead to the condition $({\cal N} -{\cal J})\times {\cal I} =3$.  
An acceptable case for three generations would be 
${\cal J} =1$, ${\cal N} =2$ and ${\cal I} =3$. In this case one
obtains
$6$ $F_L$'s, $3$ $\bar F_L$'s, $3$ $F_R$'s and $6$ $\bar F_R$'s.
In addition, there are $4$ $D$'s, $9$ $h$'s and $5$ singlets.
In order to remain with the minimal spectrum of the superpotential
of Eq.(\ref {supe}), three pairs $\bar F_ L+F_L$ should become massive
through some effective superpotential term of the form  
$<\Phi>\bar F_ LF_L$. As far as the right representations
are concerned, one pair  $\bar F_ R+F_R$ should be
interpreted as the higgses $\bar H+H$ which
break the $SU(4)$ symmetry, while the remaining two additional
pairs should become massive in the same way as the left fields.  
In a similar way, we can give superheavy
masses to any other additional representations like sextets and
doublets.

$\bullet \; {\cal J}=0$.
In this case  the condition reads ${(\cal I -\cal M)\times{\cal N}}=3$. 
We may further choose ${\cal I}=4$, ${\cal M}=1$, ${\cal N}=1$ or
${\cal I}=2$, ${\cal M}=1$, ${\cal N}=3$.
For the second case for example we obtain the following spectrum:
$6 (F_L + \bar{F}_R)$, $3 (\bar{F}_L + F_R)$, $5 (1,2,2)$ LR-higgses
9 $(15,1,1)$-higgs in the adjoint of SU(4) and 2 $(1,3,1)+(1,1,3)$
pairs accompanied by 18 neutral singlet fields. 
There are no sextet fields but usually they arise naturally
in the original spectrum of the particular string model.

The remarkable feature in this spectrum is the presence of two types
of higgses which both can break the $SU(4)$ symmetry. This fact gives
various possibilities of symmetry breaking patterns which will be
described in a future publication.

In the following we  present a string toy example\cite{alp} based 
on the Pati-Salam (PS) observable gauge symmetry and
a specific hidden interaction which can play the role of
confining gauge group, previously referred  as $SU(N)$.
 This will enable us to give a more detailed
description of the suggested approach of the previous sections.
 We will work in the free fermionic formulation of the four dimensional
superstring using free world - sheet fermions which pick up phases\cite{mb}
when parallel transported around the string. A specific choice of
phases -- consistent with the modular invariance constraints -- for
all world - sheet fermions of the left (supersymmetric) and right 
(non-sypersymmetric) sectors of the string, define a `phase' vector. 
The specific choice of the boundary conditions, will determine the
exact gauge group of the theory and the specific transformation 
properties of the preonic fields we are using. More precisely,
these vectors break simultaneously the original high string ($SO(44)$)
symmetry to the observable $PS$--gauge group and the hidden part.
In particular, the supermultiplets  of the observable gauge group
are formed from the fractionally charged states which
transform as $4$ or $\bar 4$ representations of an $SU(4)$
hidden gauge group. Although this is not a fully realistic
model, it serves as an example where the basic features
described so far in the field theory approach of the previous
section, as well as the basic ingredients of the model,
 can be found in the string spectrum of $k=1$ models.
The rank of the $SU(4)_{Hidden}$ group is smaller than the one
needed to confine the preonic matter at the conventional
GUT scale of $~10^{16}GeV$. 
In the most optimistic case, the $SU(4)$ confining scale  
is of the order $~10^{12}GeV$, i.e. $3-4$ orders smaller than the
conventional GUT scale. In models with PS - gauge symmetry 
such a low scale is not disastrous. In fact, there are no gauge bosons
mediating proton decay and the SU(4) breaking scale can be low
enough provided this is consistent with the low energy values of
gauge couplings. This is indeed the case revealed in several recent
renormalization group analyses\cite{nnn,last} of models based on 
the PS- gauge group. Note however that a realistic model should
at least have $SU(5)$ or a higher rank symmetry as a confining group.

Let us now present the basis of `phase vectors' generating the preonic
matter discussed above, at the string level. 
We denote the 18 real free fermions of the supersymmetric left-moving 
sector with $\chi^{1...6},y^{1...6},\omega^{1...6}$. For the right
moving 22 complex fermions we use the notation
$\bar {\Psi}^{1...5},\bar {\Phi}^{1...9}$ and $\bar {z}^{1...8}$.
Now a particular basis is generated by  assuming consistent\cite{mb} 
boundary conditions on the world - sheet  fermions of the two  sectors 
of the heterotic string. In our case, we assume six vectors of boundary
conditions which form a group under addition modulo 2. The 
basis is the following:
\bea
\begin{array}{lllllll}
1&=&\{\psi^{\mu},&\chi^{1...6}&,y^{1...6}&,
\omega^{1...6};
&\bar {\Psi}^{1...5}\bar {\Phi}^{1...9}\bar {z}^{1...8}\}\\

S&=&\{\psi^{\mu},&\chi^{1...6}&,0,...,0;&0,...,0&0,...,0\}\\

b_1&=&\{\psi^{\mu},&\chi^{12},&y^{3456}, 
&0,...,0;&\bar{\Psi}^{123}\bar{\Phi}^{123} 
\underbrace{\bar z^{1...8}}_{1/2}\}\\

b_2&=&\{\psi^{\mu},&\chi^{34},&y^{1256},
&0,...,0;&\bar {\Psi}^{45}\bar {\Phi}^{123}
\underbrace{\bar {\Phi}^{78}}_{1/2}
\underbrace{\bar z^{1...6}}_{-1/2}\bar z^7\}\\

b_3&=&\{00,&{0...0},&y^{1...6}, 
&\omega^{1...6};&\bar {\Phi}^{78}\}\\

\zeta&=&\{00,&{0...0},&0,...,0, 
&0,...,0;&\bar {\Phi}^{123}\bar {\Phi}^{678}\bar {z}^{12}\}
\end{array}\label{basis}
\eea
All world sheet - fermions appearing in the vectors -- except those
underlined with $\pm 1/2$ -- possess periodic boundary conditions,
while those not appearing in a particular vector are antiperiodic.
The symmetry breaks down to the following product group
\begin{eqnarray}
[SU(4)\times SU(2)_L\times SU(2)_R]_{Observable}\times
\quad\quad\quad\quad\quad\nonumber\\
\quad\quad\quad\quad\quad
[\{SU(4)^2\}_C\times SU(4)'\times SU(2)'\times
SU(2)^{''}\times U(1)^6]_{Hidden}
\label{eq:sym}
\end{eqnarray}
where the {\em Observable} and the {\em Hidden} sectors of the symmetry
are denoted with subscripts.  The particular content of the model depends on
the choice of the  specific set of the projection coefficients
 $c\left[{}^{b_i}_{b_j}\right]$. One possible choice
is   $c\left[{}^{1}_{1}\right] = c\left[{}^{\zeta}_{\zeta}\right] =-1$,
 $c\left[{}^{b_1}_{b_1}\right] =  c\left[{}^{b_2}_{b_2}\right] = -1$,
$c\left[{}^{b_i}_{\zeta}\right] = -1 $, $c\left[{}^{S}_{b_2}\right]=-1$, 
while  $c\left[{}^{i}_{j}\right]$=+1 for the remaining with $i\le j$
in the order of appearance in Eq.(\ref{basis}).
All the others are fixed from the modular invariance constraints.

The spectrum which arises is listed below, where the quantum numbers refer to
the {\em Observable} sector groups, the two $\{SU(4)^2\}_C$ groups and the 6 
$U(1)$s.
\begin{eqnarray} 
\begin{array}{llll}
b_1:& K_1^c & = & (\bar 4,1,1)(4,1)_{(0,0,1/2,1,1/4,1/4)}\\


3b_1:&\bar K_2 & = & ( 4,1,1)(\bar 4,1)_{(0,0,-1/2,-1,-1/4,-1/4)}\\

b_2:& \alpha_{1L} & = & (1,2,1)(4,1)_{(1/2,0,-1/2,-1,1/2,0)}\\


3b_2:&\bar \alpha_{2L} & = & ( 1,2,1)(\bar 4,1)
_{(-1/2,0,1/2,1,-1/2,0)}\\

2b_1+b_2:& \bar \alpha_{1R} & = & (1,1,2)(\bar 4,1)
_{(1/2,0,1/2,1,0,-1/2)}\\


2b_1+3b_2:&\alpha_{2R} & = & ( 1,1,2)( 4,1)
_{(-1/2,0,-1/2,-1,0,1/2)}\\

\end{array}
\end{eqnarray} 

An equal number of fractionally charged states transforming as $4$ or 
$\bar{4}$ under the second of the $\{SU(4)^2\}_C$, differing only in the 
U(1) factors, arises if we add the vector $\zeta$ to the above sectors.
The above fields are accompanied by singlet fields $\xi_i$, $\zeta_i$
 which arise from the Neveu-Schwarz (N-S) and $\zeta$ sectors and two
$(6,1,1)+(1,2,2)$ representations from the $2b_1+3b_2+\zeta $ sector. 
Finally there are four $ SU(4)'\times SU(2)'$ representations 
transforming as  $Q=(4,2)'$, $\bar{Q}=(\bar 4,2)'$.

All the resulting representations  of the observable sector have
double multiplicity while they transform as the $4$ or $\bar 4$ under one
of the two $SU(4)_{C}$ groups. In particular, fermion like
 condensates arise from the combinations $F_{1L}=\bar{K_2}\alpha_{1L}$
and $F_{1R}={K_1^c}{\bar\alpha}_{1R}$. Similarly one can obtain the rest 
of the spectrum presented in Table I.
Unfortunately, the number of preonic states in this toy example does not
meet the conditions put previously in order to obtain the right number
of generations and higgs fields, at least not directly at the string level.
However, in general, this fact does not exclude the possibility of 
finding a flat direction where some of the singlet fields get non - zero 
vevs and make massive the superfluous preonic fields through trilinear
couplings of the superpotential shown in the Appendix.
Thus, our string toy example, although not a realistic one, it is
deductive with respect to what one should expect from a string derived
spectrum. For example, families (and other fields) are distinguished from each
other by different $U(1)$ factors accompanying the PS symmetry. Thus the family
constructed above has quantum numbers 
$F_{1L}  =  ( 4,2,1)_{(\frac{1}{2},0,-1,-2,\frac{1}{4},-\frac{1}{4})}$ 
and
$F_{1R}  =  (\bar 4,1,2)_{(\frac{1}{2},0,1,2,\frac{1}{4},-\frac{1}{4})}$.
The appearance of U(1) symmetries is an encouraging fact as it may
generate the desired fermion mass hierarchy.

Let us finally discuss how we reach the $N=1$ supersymmetry.
The element $S$ of the above basis, with exactly 8 left movers,
plays the role of supersymmetry generator in the fermionic
construction. The subset $\{ 1, S, 1+\zeta\}$ defines an $N=4$
space - time supersymmetric model, while the introduction of
the rest of the basis vectors break successively the $N=4$
 supersymmetries to $N=2$ and $N=1$. In the class of models we
are dealing with, the confinement scale $M_C$ is assumed to
be no less than the GUT scale, i.e., the $SU(4)$ - breaking
scale. 
Thus, at the scale $M_C$ we are left  with an $N=1$ 
supergravity model based on PS - observable gauge symmetry.
Now, the only consistent way to break $N=1$ local supersymmetry
is spontaneously via the super Higgs mechanism. 
No matter what the mechanism of supersymmetry breaking is
(dynamical breaking
\cite{dsb},
gaugino condensation
\cite{gc},
coordinate dependent compactifications 
\cite{comp} 
etc) a natural hierarchy 
$m_{3/2}\ll M_{Planck}$ should be generated, 
where $m_{3/2}$  is the gravitino mass which sets the scale
of supersymmetry breaking. 
This in turn implies that there should exist a class of string 
effective supergravities where certain conditions are satisfied
like absence of fine tuning, vanishing cosmological constant
up to ${\cal O}(m_{3/2}^4)$ corrections etc. It has been
shown
\cite{FKZ}
that such conditions can be met 
particularly in the four dimensional string fermionic
constructions  we are dealing with in this work. 
In particular, supersymmetry breaking via gaugino condensation
can be shown
\cite{FKZ}
to exist in examples for superpotentials
with  non-trivial dependence on the dilaton field $S$, with 
a well behaved  positive - semi -  definite potential. 
However, as the the Hidden gauge group of this kind of models
is  rich, a dynamical supersymmetry  breaking scenario
\cite{dsb}
is quite possible here
\cite{md/af}.
In such models,  
the messengers could be states of a Hidden gauge group. 
In the present case, the $Q$, $\bar{Q}$ representations of the
$SU(4)'\times SU(2)'$ Hidden symmetry do not couple to the
ordinary matter representations, whilst they form trilinear
couplings $\Phi_iQ\bar{Q}$  with 
the gauge singlet fields $\Phi_i=\xi_i,\zeta_i$ of the N-S and
$\zeta$ sectors, in the tree level superpotential. 
Due to the presence of the massless matter
multiplets in the spectrum, the Hidden gauge group $SU(4)'$
may be strongly interacting at a relatively low scale $\Lambda$. 
Indeed, the beta function $b_4'=-12+2n_4 = -4$ in this case, while
the scale is given by
\begin{equation}
\Lambda= M_{string}
 exp\{\frac{2\pi}{b_{4}'}
(\frac{1}{\alpha_{string}}-\frac{1}{\alpha_{\Lambda}})\}
\end{equation}
Thus, for $M_{string}\sim 5\times 10^{17}GeV$ and $\alpha_{string}=1/24$,
$\alpha_{\Lambda}$ becomes $\sim 0.2$ at $\Lambda \sim 5\times 10^4 GeV$.
Now, one of the singlet fields may obtain a non-vanishing vev along
the scalar and $F$- components. As a result, gaugino masses arise at
 one loop, being directly proportional to the scale $\Lambda$ 
\begin{equation}
m_i \sim \frac{\alpha_i}{4\pi}\Lambda
\end{equation} 
This shows that the scale $\Lambda$ is of the desired order 
for a dynamical symmetry breaking scenario.

In conclusion,
in this note we have examined the possibility of obtaining
low energy effective gauge models using only representations
which arise at k=1 level of string constructions, with gauge 
symmetry  based on the $SU(4)\times SU(2)_L\times SU(2)_R$ 
Pati-Salam (PS) gauge group.  
We have argued that even though the adjoint or higher 
representations are absent from the  spectrum of such string
derived models, there are still various possibilities of 
obtaining a consistent low energy phenomenology. In particular,
in constructions based on the aforementioned PS-gauge
symmetry and k=1 Kac-Moody level, we have seen 
that a viable low energy phenomenological theory can be 
derived in the following ways:\\
i) One can use the standard PS-representations $(4,2,1)$,
 $(\bar 4,1,2)$ to accommodate the three fermion families,
while the standard gauge symmetry is obtained after the 
spontaneous breaking of the PS symmetry with a minimum
number of Higgses sitting in  $(4,1,2)$ +  $(\bar 4,1,2)$ 
representations\cite{alr}. The standard model can break with
the use of the two standard doublets found in the (1,2,2)
of the PS-symmetry.
These standard representations are accompanied by the ``exotics''
$(4,1,1)$, $(\bar 4,1,1)$ and $(1,2,1)$, $(1,1,2)$ which carry
fractional electric charges, and should form massive states
at a rather high scale to avoid phenomenological problems.\\
ii) As a second possibility, we have argued in this paper that
the above ``exotics'' may arise with non-trivial transformation
properties under a ``hidden'' gauge group with sufficient rank,
in order to confine to integral charged states at a high scale. 
It has been shown that the resulting condensates can have the
correct transformation properties to accommodate quark, lepton
and higgs fields and reproduce the model of case i). In addition,
new symmetry breaking patterns can be obtained as it is possible
now to accommodate higgs fields in the adjoint representation.
The models of case ii) are reminiscent of supersymmetric composite
models, proposed  sometime ago\cite{ven,revs,pa}.

\vspace{.5cm}
We would like to thank C. Kounnas for helpful discussions.
The work of NDT is partially supported by the C.E.C Projects SC1 -
CT91 - 0729 and CHRX - CT93 - 0319.


\end{document}